# Prokaryotic regulatory systems biology: Common principles governing the functional architectures of *Bacillus subtilis* and *Escherichia coli* unveiled by the natural decomposition approach

**Julio A. Freyre-González[1,2]§‡, Luis G. Treviño-Quintanilla[3], Ilse A. Valtierra-Gutiérrez[2], Rosa María Gutiérrez-Ríos[1] and José A. Alonso-Pavón[2]**

[1]Department of Molecular Microbiology, Institute for Biotechnology, Universidad Nacional Autónoma de México. Apdo. Postal 510-3, 62250. Cuernavaca, Morelos, México. [2]Undergraduate Program on Genomic Sciences, Center for Genomic Sciences, Universidad Nacional Autónoma de México. Av. Universidad s/n, Col. Chamilpa, 62210. Cuernavaca, Morelos, México. [3]Department of Biotechnology, Universidad Politécnica del Estado de Morelos. Blvd. Cuauhnáhuac 566, Col. Lomas del Texcal, 62550. Jiutepec, Morelos, México.

§Corresponding author:
Fax: +52(777)329-1701
E-mail: jfreyre@ccg.unam.mx (JAF-G)

‡Present address: Genome Dynamics Program, Center for Genomic Sciences, Universidad Nacional Autónoma de México. Av. Universidad s/n, Col. Chamilpa, 62210. Cuernavaca, Morelos, México.

**Abstract**
*Escherichia coli* and *Bacillus subtilis* are two of the best-studied prokaryotic model organisms. Previous analyses of their transcriptional regulatory networks have shown that they exhibit high plasticity during evolution and suggested that both converge to scale-free-like structures. Nevertheless, beyond this suggestion, no analyses have been carried out to identify the common systems-level components and principles governing these organisms. Here we show that these two phylogenetically distant organisms follow a set of common novel biologically consistent systems principles revealed by the mathematically and biologically founded natural decomposition approach. The discovered common functional architecture is a diamond-shaped, *matryoshka*-like, three-layer (coordination, processing, and integration) hierarchy exhibiting feedback, which is shaped by four systems-level components: global transcription factors (global TFs), locally autonomous modules, basal machinery and intermodular genes. The first mathematical criterion to identify global TFs, the κ-value, was reassessed on *B. subtilis* and confirmed its high predictive power by identifying all the previously reported, plus three potential, master regulators and eight sigma factors. The functional conserved cores of modules, basal cell machinery, and a set of non-orthologous common physiological global responses were identified via both orthologous genes and non-orthologous conserved functions. This study reveals novel common systems principles maintained between two phylogenetically distant organisms and provides a comparison of their lifestyle adaptations. Our results shed new light on the systems-level principles and the fundamental functions required by bacteria to sustain life.

# 1  Introduction

Over the last two decades, high-throughput methods and literature curation efforts have provided a myriad of biological information, now available in specialized databases. Nowadays, all this data requires novel approaches in order to integrate a biologically coherent systems view of cell control. Graph theory has provided a theoretical framework to globally analyze sets of molecular interactions, looking for universal laws governing them (Barabasi, 2009; Barabasi and Oltvai, 2004; Milo et al., 2002). In particular, different studies have proposed methodologies to infer the modules, hierarchy, or both shaping a transcriptional regulatory network (TRN) (Balazsi et al., 2005; Bhardwaj et al., 2010b; Janga et al., 2009; Ma et al., 2004a; Ma et al., 2004b; Resendis-Antonio et al., 2005; Rodriguez-Caso et al., 2009; Yu and Gerstein, 2006). Although analyses using these methods have reported good results, they are not natural and have revealed some inconveniencies: 1) some of them disregards genes not encoding transcription factors (TFs) (Janga et al., 2009; Resendis-Antonio et al., 2005; Rodriguez-Caso et al., 2009), 2) others rely on certain parameters that, when modified, can generate different modules (Resendis-Antonio et al., 2005), 3) even others cluster known pleiotropic TFs into a module (Resendis-Antonio et al., 2005; Rodriguez-Caso et al., 2009) or place them in lower hierarchical layers (Bhardwaj et al., 2010b; Ma et al., 2004a; Ma et al., 2004b; Yu and Gerstein, 2006), and 4) some others have proven to be inadequate for networks comprising feedback loops (FBLs) or feedforward motifs (Balazsi et al., 2005; Bhardwaj et al., 2010b; Ma et al., 2004a; Ma et al., 2004b; Yu and Gerstein, 2006), two topological structures relevant to the organization and dynamics of TRNs (Barabasi and Oltvai, 2004; Milo et al., 2002; Smits et al., 2006; Thomas, 1998).

Recently, these issues have been tackled in order to reveal the functional architecture and systems-level components of the TRN of *Escherichia coli* by developing and applying the natural decomposition approach (NDA) (Freyre-Gonzalez et al., 2008). This biologically motivated approach mathematically derives the architecture and systems-level components from the whole structure of a given TRN and is founded on two pertinent premises: 1) A module is a set of genes cooperating to carry out a particular physiological function (Hartwell et al., 1999), thus conferring different phenotypic traits to the cell. 2) Given the pleiotropic effects of global regulators (Gottesman, 1984), they must not belong to modules but rather coordinate them in response to general-interest environmental cues (Freyre-Gonzalez et al., 2008). According to this approach, all genes in the *E. coli* TRN are predicted to belong to one out of four possible systems-level components. These components interrelate in a non-pyramidal hierarchy shaping the *E. coli* functional architecture as follows: 1) global TFs are responsible for coordinating both the 2) basal cell machinery, composed of strictly globally regulated genes (SGRGs), and 3) locally autonomous modules (shaped by modular genes), whereas 4) intermodular genes integrate, at promoter level, physiologically disparate module responses eliciting a combinatorial processing of environmental cues.

There exist dramatic differences in how gram-positive and gram-negative bacteria regulate gene expression even globally (Lozada-Chavez et al., 2008; Price et al., 2007; Sonenshein et al., 2002). In fact, different works have studied the evolution of TFs, showing that regulatory networks exhibit high plasticity during evolution (Janga and Perez-Rueda, 2009; Lozada-Chavez et al., 2006; Madan Babu et al., 2006; Price et al., 2007). In particular, a study by Babu *et al.* analyzed the evolution of local and global network structures in *E. coli* and *Bacillus subtilis* (Madan Babu et al., 2006). All these studies have shown that TRNs exhibit high plasticity during evolution converging to scale-free-like structures. Nevertheless, some aspects still remain unknown, such as the exact nature of these structures (*e.g.*, whether they are true hierarchical-modular networks or other type of scale-free structures), their systems-level components (*e.g.*, modules and intermodular genes) and the systems principles governing them. In order to address these issues, in this study we applied the NDA to reveal the systems principles and systems-level components governing the TRN of the soil bacterium *B. subtilis*, which is currently the best-characterized TRN for a gram-positive model organism (Sierro et al., 2008). Next, we integrate these components through inferring the hierarchy controlling the *B. subtilis* TRN. We then compared the inferred hierarchy and the systems-level components distribution to the ones previously reported that governs the *E. coli* TRN (Freyre-Gonzalez et al., 2008), which were systematically contrasted with experimental data collected in RegulonDB (Gama-Castro et al., 2010). Our analyses revealed a set of common system-level principles governing the organization of the two, otherwise different as has been reported previously (Price et al., 2007; Vazquez et al., 2009), TRNs. Finally, in a complementary part of our study, we explored the functional conservation between *B. subtilis* and *E. coli* of the different systems-level components via both orthologous genes and non-orthologous conserved functions. Our results reveal a set of common physiological responses and the systems responsible for them, what we defined as the conserved cores of the basal cell machinery and modules. Some of these systems exhibit a statistical significant overlap via orthologs and, therefore, have possibly been conserved from a common ancestor. Whereas some other systems conserve the same function in both organisms but are not composed by any orthologs, thus suggesting that they has been independently rediscovered anytime after divergence from a common ancestor.

## 2 Materials and methods

### 2.1 Data extraction and reconstruction of the TRN

The *B. subtilis* TRN was reconstructed using data from the DBTBS database (Sierro et al., 2008). The database was provided as a XML file by the DBTBS team. Next, we parsed this file to extract the regulatory interactions (the resulting data set is available as Additional file 5). Structural genes, genes encoding sigma factors and regulatory proteins were included in our graphic model. In order to avoid duplicated interactions, heteromeric TFs were represented as only one node provided that there is no evidence indicating that any of the subunits have a regulatory activity *per se*.

### 2.2 Natural decomposition approach

The NDA defines criteria to identify each systems-level component in a TRN, thus revealing its functional architecture. The NDA proceeds as follows: 1) Global TFs are identified by computing the κ-value. For each node in the TRN, relative connectivity (as a fraction of maximum connectivity, $k_{max}$) and clustering coefficient are computed. Next, the $C(k)$ distribution is obtained using least-squares fitting. Finally, given $C(k) = \gamma k^{-\alpha}$, the κ-value is calculated using the formula $\kappa = \sqrt[\alpha+1]{\alpha\gamma} \cdot k_{max}$, which is the solution to the equation $dC(k)/dk = -1$. Global TFs are those nodes exhibiting connectivity greater than κ. 2) Global TFs and their links are removed from the network, thus naturally revealing the modules (isolated islands composed of interconnected nodes) and the SGRGs (single disconnected nodes). 3) Intermodular genes are identified by analyzing the megamodule. The megamodule is isolated and all structural genes (non-TF-encoding genes) are removed, disaggregating it into isolated islands. Next, each island is identified as a pre-submodule. Finally, all the previously removed structural genes and their interactions are added to the network according to the following rule: if a structural gene G is regulated only by TFs belonging to submodule M, then gene G is added to submodule M. On the contrary, if gene G is regulated by TFs belonging to two or more submodules, then gene G is classified as an intermodular gene.

## 2.3 Manual annotation of identified modules

We annotate each module as follows: First, we compiled a list of biochemical functions and cellular processes in which the genes composing each module were involved. We obtained this information from the DBTBS database (Sierro et al., 2008) and a review of pertinent literature. Then, this information was analyzed taking into account the regulatory and physiological context by using expert biological knowledge to make human inferences about the physiological function of each module.

## 2.4 Computational annotation of identified modules

Each gene was annotated with its corresponding functional class according to the DBTBS database (Sierro et al., 2008). Next, *p*-values, as a measure of randomness in functional class distributions through identified modules, were computed based on the hypergeometric distribution. Let *N* be the total number of genes in the TRN and *A* the number of these genes with a particular *F* annotation; the *p*-value is defined as the probability of observing, at least, *x* genes with an *F* annotation in a module with *n* genes. This *p*-value is computed as follows:

$$p\text{-value} = \sum_{i=x}^{\min(n,A)} \frac{\binom{A}{i}\binom{N-A}{n-i}}{\binom{N}{n}}.$$

Thus, for each module, the *p*-value of each functional assignment present in the module was computed. The functional assignment of the module was the one showing the lowest *p*-value, if and only if it was less than 0.05.

## 2.5 Algorithm for FBLs enumeration

In order to enumerate all the FBLs in the *B. sublitis* TRN, we applied the algorithm proposed by Freyre-González *et al.* (Freyre-Gonzalez et al., 2008). This algorithm optimizes the FBLs identification, which is a computationally intractable problem, by reducing the search space through excluding nodes not belonging to any FBL. First, computing the transitive closure matrix of the TRN identifies the set of nodes participating in FBLs. Next, FBLs are enumerated by depth-first searches that are initiated for every node participating in circular interactions. Finally, FBL isomorphisms are removed.

### 2.6 TRN-wide orthologs identification

To evaluate the conservation of systems-level components between the *B. subtilis* and *E. coli* we computed TRN-wide orthologs. First, 1208 genome-wide orthologs between *B. subtilis* 168 and *E. coli* K-12 MG1655 were identified using the Comparative Analysis tool available via the BioCyc database version 13.0 (Karp et al., 2005). This tool defines two proteins A and B as orthologs if A and B are bi-directional best BLAST hits of one another, meaning that B is the best BLAST hit of A within the genome of B, and A is the best BLAST hit of B within the genome of A, which is an optimal and straightforward method to identify orthologous pairs on a genome-wide scale. Closest hits are defined as ones having the smallest E-value (cut-off 0.001). Ties are broke by comparing the length of the different hits' high scoring pairs. Then, we defined a TRN-wide ortholog as a genome-wide orthologous gene pair such that each gene is included in its corresponding TRN. We identified 312 TRN-wide orthologs.

### 2.7 Module conservation index

In order to quantify the conservation level of modules sharing TRN-wide orthologs between *B. subtilis* and *E. coli*, we defined the module conservation index as follows: given any two modules $M_i$ and $M_j$, the module conservation index is computed as:

$$C_{i,j} = \frac{|M_i \cap M_j|}{\min(|M_i|,|M_j|)},$$

where $|M_i|$ denotes the number of elements in $M_i$, and $M_i \cap M_j$ represents the set of TRN-wide orthologs shared between $M_i$ and $M_j$. Module conservation index tells how conserved is a given pair of modules but tells nothing about differences arising from the evolutionary expansion/contraction in the number of genes composing each module. In order to complete the picture, we defined the module size log ratio as:

$$S_{i,j} = \log\left(\frac{|M_i|}{|M_j|}\right).$$

### 2.8 Statistical significance of module conservation indices

For each pair of modules $M_i$ and $M_j$ with module conservation index greater than zero, we computed the hypergeometric probability of the overlap because of chance. Let $N$ be the total number of TRN-wide orthologs in modules in both organisms, $n$ and $m$ be the number of orthologs in modules $M_i$ and $M_j$, respectively, and $k$ be the number of orthologs shared by modules $M_i$ and $M_j$ ($|M_i \cap M_j|$). The *p*-value was defined as:

$$p\text{-value} = \sum_{i=k}^{\min(n,m)} \frac{\binom{m}{i}\binom{N-m}{n-i}}{\binom{N}{n}}.$$

## 3 Results and discussion

We reconstructed the *B. subtilis* TRN using data available in the DBTBS database (Sierro et al., 2008) (see Materials and methods). In our model, nodes and arcs (directed links) represent genes and regulatory interactions, respectively. The reconstructed TRN comprised 1679 nodes (~40% of the total genes in the genome) and 3019 arcs between them. Neglecting 73 autoregulatory interactions and directionality, we found that connectivity and clustering coefficient distributions follow power laws (Figure S1a,b in Additional file 1), which suggests that the *B. subtilis* TRN exhibits hierarchical-modularity.

### 3.1 *B. subtilis* global TFs are accurately predicted by the κ-value

The NDA defines an equilibrium point (κ-value) between two apparently contradictory behaviors occurring in hierarchical-modular networks: hubness and modularity (Freyre-Gonzalez et al., 2008). Both properties depend on the particular organization of a given network and can be quantified by computing the connectivity and clustering coefficient, respectively, in a per gene basis. In hierarchical-modular networks, it has been shown that modularity decreases as hubness increases (Barabasi and Oltvai, 2004; Freyre-Gonzalez et al., 2008). Then, the κ-value is defined as the connectivity value for which the variation of the clustering coefficient (modularity) equals the variation of connectivity (hubness) but with the opposite sign ($dC(k)/dk = -1$) (Freyre-Gonzalez et al., 2008). This is a network-depending parameter that allows the identification of global TFs as those nodes with connectivity greater than κ (see Materials and methods). We identified 19 global TFs by computing the κ-value for the *B. subtilis* TRN (Figure 1): SigA, SigE, CcpA, SigB, SigK, SigG, AbrB, SigD, ComK, TnrA, SigW, SpoIIID, LexA, PhoP, CodY, Fur, SigF, ResD and Spo0A (see section 3.5 for further discussion of the physiological role that global TFs play in the hierarchy controlling the TRN). These TFs comprise 8 out of the 15 sigma factors included in the analyzed TRN. Except for SpoIIID, LexA and ResD, all of these TFs have been jointly described as global TFs (Haldenwang, 1995; Madan Babu et al., 2006; Moreno-Campuzano et al., 2006). Nevertheless, SpoIIID (Wang et al., 2007), LexA (Groban et al., 2005; Wojciechowski et al., 1991) and ResD (Geng et al., 2007) have also been individually reported as global TFs. Each of these TFs responds to general environmental cues and controls a large number of genes in the cell. The accurate predictive power of the mathematically defined κ-value to identify global TFs in distantly related bacteria strongly suggests this is a fundamental biological parameter governing the organization of prokaryotic TRNs.

## 3.2 Common systems-level principles and components maintained between the functional architectures of *B. subtilis* and *E. coli*

According to the second step in the NDA, we removed all global TFs (see Materials and methods). This revealed 850 SGRGs, 69 independent modules, and one megamodule. After isolating the megamodule, we found that the removal of all non-TF-encoding genes disintegrated the megamodule into several isolated islands, as previously was also found for *E. coli* (Freyre-Gonzalez et al., 2008). This suggests that some non-TF-encoding genes are coregulated by TFs belonging to different submodules thus holding the megamodule together. Then, following the third step in the NDA, a given non-TF-encoding gene was classified as intermodular gene if the TFs regulating it belong to different modules (see Materials and methods). We identified 42 intermodular genes, organized into 24 operons (see Additional file 2 for a detailed biological discussion). A comparison between these results and the previously found for *E. coli* revealed that the *B. subtilis* TRN is composed by the same four systems-level components appearing in approximately the same relative amount (Table 1). These system-level components shape an architecture comprising three executive layers: coordination (controlled by global TFs), processing (comprising locally autonomous modules and SGRGs) and integration (composed of intermodular genes). As for *E. coli* (Freyre-Gonzalez et al., 2008) and contrary to previously proposed pyramidal hierarchies (Balazsi et al., 2005; Bhardwaj et al., 2010b; Janga et al., 2009; Ma et al., 2004a; Ma et al., 2004b; Martinez-Antonio and Collado-Vides, 2003; Rodriguez-Caso et al., 2009; Yan et al., 2010; Yu and Gerstein, 2006), a diamond-shaped hierarchy is evidenced by the fact that the number of genes in the, middle, processing layer is much greater than those in the, lower, integrative layer (Table 1) as a consequence of the integration, at promoter level, of signals coming from different modules that process disparate physiological conditions (see Additional file 2). An example of this organization is the *gpr* gene that encodes a protease that initiates degradation of small, acid-soluble spore proteins during the first minutes of germination. The expression of this gene is combinatorially activated by RsfA (module 2.r9; initial phase of sporulation), repressed by SpoVT (module 2.r11; final phase of sporulation, and germination) and also governed by two global TFs, $\sigma^F$ and $\sigma^G$. The expression of *rsfA* is orchestrated by $\sigma^F$ and $\sigma^G$, and auto-repression; while *spoVT* is auto-repressed and transcribed by $\sigma^G$.

## 3.3 Biological relevance of the *B. subtilis* modules, SGRGs and intermodular genes

The biological relevance of the identified modules was evaluated by two independent blind analyses: 1) One of us (LGT-Q) used biological knowledge to annotate each identified module. 2) Other of us (JAF-G) used hypergeometric probabilities to carry out a blind-automated annotation based on statistically significant functional classes ($p$-value < 0.05) (see Materials and methods). The blind-automated annotation found that 80% of the modules could be annotated at the given significance level. Both analyses produced similar results (Table S1 in Additional file 3). However, the manual analysis added subtle details that were not evident in the automated analysis because of incomplete annotations in the used functional classification. We noted that *B. subtilis* modules mainly compose systems involved in carbon source metabolism (27%), sporulation and germination (10%), amino acids metabolism (9%) and protein biosynthesis (9%) (Figure 2a). Interestingly, while *B. subtilis* and *E. coli* modules are mainly involved in carbon sources metabolism (27% and 38%, respectively), the percentage of modules involved in stress response decreased from 11% in *E. coli* to 7% in *B. subtilis* (Figure 2a and b), which is consistent with the fact that both sporulation and a global robust stress response provide an efficient alternative to cope with different stresses.

Proteins encoded by *B. subtilis* intermodular genes are mainly responsible for extracytoplasmic functions and amino acids metabolism, in contrast to *E. coli* where a common function was not evident (Freyre-Gonzalez et al., 2008). This suggests that in *B. subtilis* these functions depend on the integration of different physiological responses (see Additional file 2). Remarkably, several intermodular genes encode enzymes that are key for the biotechnological production of extracellular proteins and metabolic intermediaries. For example, *sacB* encodes an extracellular levansucrase (Pohl and Harwood, 2010), BdbA and BdbB are thiol-sulfide oxidoreductases involved in the efficient formation of disulfide bonds in proteins (Kouwen and van Dijl, 2009), and AprE is an extracellular serine alkaline endoprotease better known as subtilisin E (Nijland and Kuipers, 2008) (see Additional file 2). Besides, as we previously found in *E. coli* (Freyre-Gonzalez et al., 2008), we observed that the identified SGRGs mostly comprise elements of the basal cell machinery (*e.g.*, tRNAs and their charging enzymes, DNA and RNA polymerases, ribosomal elements, DNA repair/packing/segregation/methylation). We also noted that roughly one-third of the SGRGs comprise poorly studied systems (*i.e.*, genes annotated as "unknown function").

### 3.4 The *B. subtilis* TRN, as its *E. coli* counterpart, is also non-acyclic

After decomposing the TRN into its systems-level components, we focused our efforts in integrating these components in order to get a global image of the hierarchy governing the *B. subtilis* TRN. A problem to infer any hierarchy is the possible presence of FBLs. They are circular regulatory interactions necessary to give rise to different cellular behaviors, such as homeostasis and differentiation (Kaern et al., 2005; Smits et al., 2006; Thomas, 1998; Thomas and Kaufman, 2001), but their sole presence poses a paradox. Given the circular nature of their interactions, what nodes should be placed in a higher hierarchical layer? Some studies have resolved this paradox by claiming that TRNs are acyclic (Balazsi et al., 2005; Ma et al., 2004a); whereas other has argued that, given that genes participating in a FBL are encoded in the same operon, they can be placed in the same hierarchical layer (Ma et al., 2004b). This same-layer strategy has been applied in different studies (Bhardwaj et al., 2010a; Bhardwaj et al., 2010b; Ma et al., 2004b; Yu and Gerstein, 2006). In contrast, in a previous study, it has been shown that the *E. coli* TRN is non-acyclic, and FBLs are mostly composed by genes not encoded in the same operon, bridging different hierarchical layers (Freyre-Gonzalez et al., 2008). Neglecting the presence of FBLs can lead to biologically incorrect hierarchies placing pleiotropic TFs in lower hierarchical layers. This controversy, and its importance to infer a biologically sound hierarchy, motivated us to explore the systems-level presence of FBLs in the *B. subtilis* TRN. We enumerate all FBLs by using a previously reported algorithm that deals with the computational intractability through reducing the search space (Freyre-Gonzalez et al., 2008) (see Materials and methods).

We found 16 FBLs, none comprising genes encoded in the same operon (Table 2). The latter provides further support to the fact that it is not safe to apply the same-level strategy to deal with FBLs. A dual FBL comprising *phoP* and *resD* was found because of PhoP might activate or repress *resD* expression (Eldakak and Hulett, 2007). Both FBLs were enumerated because their dynamic behaviors are completely different. Every identified FBL comprises at least one global TF, therefore showing that FBLs are not inside modules. This is consistent with previous results showing that 80% of the *E. coli* FBLs are not in modules. On the contrary, 68.8% (11/16) of the *B. subtilis* FBLs comprise both global and modular TFs, while the remaining 31.2% (5/16) only involve global TFs. This implies that a TRN could include a democratic structure where some modular TFs reconfigure the global transcriptional machinery, through affecting the expression of some global TFs, in response to specific conditions. In this context, an interesting FBL is the one shaped by GerE (sporulation-specific transcriptional factor, a modular TF) and SigK (sigma factor in the mother cell following engulfment, a global TF). An experimental study has provided evidence showing that this FBL plays a key role in enhancing the robustness of the mother cell network and optimizing the expression of target genes (Wang et al., 2007).

While FBLs are not statistical enriched in the TRN, the low number of genes involved in them, in both *B. subtilis* and *E. coli* TRNs (16 and 24 genes respectively), is in agreement with a recent study proposing that carefully placing FBLs is a design principle in complex networks that provides dynamical stability (Ma'ayan et al., 2008). We also found that 62.5% (10/16) are positive, which potentially could give rise to bistability and thus differentiation, in contrast to only 35% (7/20) in *E. coli*. In fact, it has been proposed that FBLs coupled with noise in gene expression are key in the formation of *B. subtilis* bistable populations with respect to at least the Spo0A, SigD and ComK global TFs (Aguilar et al., 2007). Remarkably, all of these TFs are exclusively involved in positive FBLs (Table 2).

## 3.5 Common themes and differences in the functional hierarchies governing *B. subtilis* and *E. coli*

We solve the paradox arising from the presence of FBLs and inferred the hierarchical organization governing the *B. subtilis* TRN by applying a previously proposed algorithm that assign genes to layers according to their theoretical pleiotropy (Freyre-Gonzalez et al., 2008). The revealed structure is composed of modules nested into others like a set of Russian nesting dolls or *matryoshka* (Figure 3a), supporting the previously suggested self-similar nature of complex biological networks (Song et al., 2005) and showing that this is not a myth as it was asserted recently (Lima-Mendez and van Helden, 2009). We found nine mid-level metamodules (*i.e.*, higher-level modules comprising other modules) in the *B. subtilis* TRN (Figure 3a). Spo0A is the central element in the hierarchy, globally governing six out of the nine mid-level metamodules. It represses three mid-level metamodules, while activating other three. This divides the hierarchy into three regions: 1) Vegetative growth, which comprises six mid-level metamodules: iron transport, motility and chemotaxis, extracytoplasmic and antibiotics-induced stress, SOS response, nitrogen metabolism, and competence (Chastanet and Losick, 2011). 2) Sporulation, which is composed of two Spo0A-activated mid-level metamodules: forespore control (coordinated by SigG and SigF) and mother cell control (coordinated by SigE, SpoIIID and SigK) (de Hoon et al., 2010). 3) A common region containing a Spo0A-activated mid-level metamodule related to phosphate metabolism and respiration, which is also controlled by CcpA, the master regulator involved in carbon source catabolism. This common region suggests a link between sporulation ability and phosphate metabolism (Eldakak and Hulett, 2007).

Five out of the nine mid-level metamodules found in *B. subtilis* overlap in terms of their physiological functions with metamodules previously identified in *E. coli* (Freyre-Gonzalez et al., 2008) (Figure 3a and b) conforming a set of common physiological global responses, these are (*B. subtilis-E. coli* TF): energy level sensing (CcpA-CRP), iron transport (Fur-Fur), motility (SigD-FlhDC), extracytoplasmic stress (SigW-RpoE), nitrogen metabolism (TnrA-RpoN), and respiration (ResD-FNR). Conversely, an interesting difference between both hierarchies is that while carbon foraging, controlled by catabolite repression, is the main physiological process globally governing the *E. coli* TRN (CRP) (Freyre-Gonzalez et al., 2008), it plays a secondary role in *B. subtilis* (CcpA) given its ability to sporulate in order to survive under adverse conditions.

### 3.6 Conservation level of the systems-level components ranges from moderate to poor and reveals the conserved core of the basal cell machinery

Given that the functional architecture is conserved, we assessed the conservation level of the four identified systems-level components. We computed TRN-wide orthologs (see Materials and methods) and mapped each ortholog to its corresponding systems-level component (Figure 4). Results showed that the most conserved systems-level component is the set of modular genes, which roughly comprise one-third of the TRN-wide orthologs. The set of SGRGs is the second most conserved systems-level component, comprising about one-sixth of the TRN-wide orthologs. Global TFs and intermodular genes are the least conserved systems-level components. Some SGRGs in an organism are orthologous to modular genes in the other. In particular, 73 (23.40%) *B. subtilis* SGRGs are distributed into 29 *E. coli* modules, whereas 34 (10.90%) *E. coli* SGRGs are distributed into 15 *B. subtilis* modules. There are three possibilities to explain this: 1) these SGRGs actually belong to modules and further research is needed to identify the missing regulatory interactions, 2) the modular TFs controlling these genes are not identified as global because of TRN incompleteness, or 3) the physiological process involving these SGRGs are so important in an organism relative to the other that they need to be strictly globally coordinated.

An analysis of the functional classes of orthologous and non-orthologous SGRGs revealed both have common functions: energy metabolism; flagellar components and motility; cell division; DNA replication, modification and repair; protein modification, folding and secretion; and transport of some cofactors and amino acids. Ortholous SGRGs additionally exhibit some exclusive functions: ribosomal proteins, amino acid metabolism, and peptidoglycan biosynthesis. Whereas non-orthologous SGRGs also show other functions: general transport systems, stress adaptation, and phage-related functions. Therefore, SGRGs shape the conserved core of the basal machinery of replication, transcription and translation of the cell. This also provides further evidence proving that the set of SGRGs is not an artifact of the NDA.

Intermodular genes are poorly conserved, an analysis of the functions of orthologous and non-orthologous intermodular genes showed that only two functions are conserved: cell division proteins (encoded by *ftsAZ*) involved in the formation of the septum, and detoxification by reducing nitrites into ammonia (mediated by a nitrite reductase encoded by *nasDEF*/*nirBDC-cysG*). The only orthologous intermodular genes are *ftsA*, *ftsZ* and *nasD*/*nirB*. *E. coli* intermodular *nirC* is orthologous to *B. subtilis ywcJ*, but the latter belongs to module 2.r12. On the other hand, *nasE*/*nirD* and *nasF*/*cysG* does not have any orthologs, but both conserve their function: the first pair encodes a nitrite reductase subunit, and the latter encodes a uroporphyrin-III C-methyltransferase responsible for the synthesis of the siroheme prosthetic group required by nitrite reductase. All these results highlight that the processing layer (modules and SGRGs) is more conserved that the coordination and integration layers. This suggests that allowing reconfiguring the combinatorial control coordinating and integrating the locally independent modular responses increase the TRN evolvability by exploring alternative control designs that could provide a better adaptive response to environmental changes.

### 3.7 A module conservation analysis reveals the conserved modular cores

In order to evaluate the conservation via orthologs between the *B. subtilis* modules found in this study and the ones previously found in *E. coli* (Freyre-Gonzalez et al., 2008), we defined and computed the module conservation index and the module size log ratio between pairs of modules (see Materials and methods). The evolutionary conservation index between modules ranges from moderate (0.67) to poor (0.01) (Figure 5 and Table S2 in Additional file 4). The module size log ratio proves that during evolution some modules, despite conserving their function, expand or contract by gaining or losing genes. We found that more than one-third of the *B. subtilis* (38/90 = 42%) and *E. coli* (37/100 = 37%) modules have a conservation index greater than zero between them, totalizing 57 pairs of hypothetical conserved modules. We validated these results statistically and biologically through 1) computing the statistical significance of module conservation from the hypergeometric probability of observing at least that many shared orthologs by chance ($p$-value < 0.05) (see Materials and methods), and 2) a functional analysis of every pair of hypothetical conserved modules, taking into account their module annotations, functional classes and biological context (Table S2 in Additional file 4). Results revealed that more than one-half (35/57 = 61%) of all pairs of hypothetical conserved modules have equivalent functions although their conservation index is moderate (0.67-0.10) (Figure 5 and Table S2 in Additional file 4). Remarkably, some of these conserved physiological functions have been recently reported as cellular essential functions to sustain life in bacteria (Azuma and Ota, 2009). Therefore, our results prove that roughly one-third of the *B. subtilis* (29/90 = 32%) and *E. coli* (29/100 = 29%) modules are functionally conserved via orthologs.

We then analyzed the functions of modules composed by non-orthologous genes and found several modular conserved functions (see Table 3). Non-conserved functions between these bacteria are related to *E. coli* modules enabling the catabolism of alternative carbon sources that do not exist in *B. subtilis*, and modules responsible for metabolize different amino acids in each bacteria. Both analyses reveal a set of common functions composing the conserved core of these modules. This highlights that module function conservation does not necessarily imply gene conservation, suggesting that convergent evolutions plays an important role in TRNs evolution. Finally, our results bring together a set of conserved genes involved in fundamental cellular functions, thus providing novel insights into the key cellular systems to sustain life.

## 4 Conclusions

Recently, network biology has been criticized for being based on myths, suggesting that the search for global principles is worthless (Lima-Mendez and van Helden, 2009). However, our results compel us to reassess this statement. The NDA provides a useful biological-mathematical framework to study the systems principles governing the organization of TRNs. By using this approach, we have unveiled the common systems principles governing the TRNs of two biotechnologically relevant prokaryotes, *B. subtilis* (this study) and *E. coli* (Freyre-Gonzalez et al., 2008), and we have provided a comparison of their lifestyle adaptations. The large evolutionary distance between these organisms suggests the possible universality among prokaryotes of the discovered common systems principles. Besides, these principles directly impact the cell modeling efforts by enabling the identification and isolation of system components and subsystems, thus reducing the complexity of the model and also impacting the algorithmic complexity of the simulation. Our results also show that orthologous genes do not explain the conservation of the fundamental cellular systems (modules) required to sustain life in bacteria, suggesting that nature is constantly searching for alternative solutions. Finally, we provide a large set of experimentally testable hypothesis ranging from novel FBLs to intermodular genes, and conserved functional cores, which we hope will contribute to inspire new experimental research in the systems biology field.

## Acknowledgments

We thank Patricia Romero for technical support. We also thank two anonymous reviewers for helpful comments and suggestions. This work was partially supported by grants 68992 from CONACyT-Salud and 60127 from CONACyT to Enrique Merino.

## Competing interests

The authors declare that they have no competing interests.

## References

Aguilar, C., Vlamakis, H., Losick, R., Kolter, R., 2007. Thinking about *Bacillus subtilis* as a multicellular organism. Curr Opin Microbiol 10, 638-643.

Azuma, Y., Ota, M., 2009. An evaluation of minimal cellular functions to sustain a bacterial cell. BMC Syst Biol 3, 111.

Balazsi, G., Barabasi, A.L., Oltvai, Z.N., 2005. Topological units of environmental signal processing in the transcriptional regulatory network of *Escherichia coli*. Proc Natl Acad Sci U S A 102, 7841-7846.

Barabasi, A.L., 2009. Scale-free networks: a decade and beyond. Science 325, 412-413.

Barabasi, A.L., Oltvai, Z.N., 2004. Network biology: understanding the cell's functional organization. Nat Rev Genet 5, 101-113.

Bhardwaj, N., Kim, P.M., Gerstein, M.B., 2010a. Rewiring of transcriptional regulatory networks: hierarchy, rather than connectivity, better reflects the importance of regulators. Sci Signal 3, ra79.

Bhardwaj, N., Yan, K.K., Gerstein, M.B., 2010b. Analysis of diverse regulatory networks in a hierarchical context shows consistent tendencies for collaboration in the middle levels. Proc Natl Acad Sci U S A 107, 6841-6846.

Chastanet, A., Losick, R., 2011. Just-in-time control of Spo0A synthesis in *Bacillus subtilis* by multiple regulatory mechanisms. J Bacteriol 193, 6366-6374.

de Hoon, M.J., Eichenberger, P., Vitkup, D., 2010. Hierarchical evolution of the bacterial sporulation network. Curr Biol 20, R735-745.

Eldakak, A., Hulett, F.M., 2007. Cys303 in the histidine kinase PhoR is crucial for the phosphotransfer reaction in the PhoPR two-component system in *Bacillus subtilis*. J Bacteriol 189, 410-421.

Freyre-Gonzalez, J.A., Alonso-Pavon, J.A., Trevino-Quintanilla, L.G., Collado-Vides, J., 2008. Functional architecture of *Escherichia coli*: new insights provided by a natural decomposition approach. Genome Biol 9, R154.

Gama-Castro, S., Salgado, H., Peralta-Gil, M., Santos-Zavaleta, A., Muniz-Rascado, L., Solano-Lira, H., Jimenez-Jacinto, V., Weiss, V., Garcia-Sotelo, J.S., Lopez-Fuentes, A., Porron-Sotelo, L., Alquicira-Hernandez, S., Medina-Rivera, A., Martinez-Flores, I., Alquicira-Hernandez, K., Martinez-Adame, R., Bonavides-Martinez, C., Miranda-Rios, J., Huerta, A.M., Mendoza-Vargas, A., Collado-Torres, L., Taboada, B., Vega-Alvarado, L., Olvera, M., Olvera, L., Grande, R., Morett, E., Collado-Vides, J., 2010. RegulonDB version 7.0: transcriptional regulation of *Escherichia coli* K-12 integrated within genetic sensory response units (Gensor Units). Nucleic Acids Res 39, D98-105.

Geng, H., Zuber, P., Nakano, M.M., 2007. Regulation of respiratory genes by ResD-ResE signal transduction system in *Bacillus subtilis*. Methods Enzymol 422, 448-464.

Gottesman, S., 1984. Bacterial regulation: global regulatory networks. Annu Rev Genet 18, 415-441.

Groban, E.S., Johnson, M.B., Banky, P., Burnett, P.G., Calderon, G.L., Dwyer, E.C., Fuller, S.N., Gebre, B., King, L.M., Sheren, I.N., Von Mutius, L.D., O'Gara, T.M., Lovett, C.M., 2005. Binding of the *Bacillus subtilis* LexA protein to the SOS operator. Nucleic Acids Res 33, 6287-6295.

Haldenwang, W.G., 1995. The sigma factors of *Bacillus subtilis*. Microbiol Rev 59, 1-30.

Hartwell, L.H., Hopfield, J.J., Leibler, S., Murray, A.W., 1999. From molecular to modular cell biology. Nature 402, C47-52.

Janga, S.C., Perez-Rueda, E., 2009. Plasticity of transcriptional machinery in bacteria is increased by the repertoire of regulatory families. Comput Biol Chem 33, 261-268.

Janga, S.C., Salgado, H., Martinez-Antonio, A., 2009. Transcriptional regulation shapes the organization of genes on bacterial chromosomes. Nucleic Acids Res 37, 3680-3688.

Kaern, M., Elston, T.C., Blake, W.J., Collins, J.J., 2005. Stochasticity in gene expression: from theories to phenotypes. Nat Rev Genet 6, 451-464.

Karp, P.D., Ouzounis, C.A., Moore-Kochlacs, C., Goldovsky, L., Kaipa, P., Ahren, D., Tsoka, S., Darzentas, N., Kunin, V., Lopez-Bigas, N., 2005. Expansion of the BioCyc collection of pathway/genome databases to 160 genomes. Nucleic Acids Res 33, 6083-6089.

Kouwen, T.R., van Dijl, J.M., 2009. Applications of thiol-disulfide oxidoreductases for optimized in vivo production of functionally active proteins in *Bacillus*. Appl Microbiol Biotechnol 85, 45-52.

Lima-Mendez, G., van Helden, J., 2009. The powerful law of the power law and other myths in network biology. Mol Biosyst 5, 1482-1493.

Lozada-Chavez, I., Angarica, V.E., Collado-Vides, J., Contreras-Moreira, B., 2008. The role of DNA-binding specificity in the evolution of bacterial regulatory networks. J Mol Biol 379, 627-643.

Lozada-Chavez, I., Janga, S.C., Collado-Vides, J., 2006. Bacterial regulatory networks are extremely flexible in evolution. Nucleic Acids Res 34, 3434-3445.

Ma'ayan, A., Cecchi, G.A., Wagner, J., Rao, A.R., Iyengar, R., Stolovitzky, G., 2008. Ordered cyclic motifs contribute to dynamic stability in biological and engineered networks. Proc Natl Acad Sci U S A 105, 19235-19240.

Ma, H.W., Buer, J., Zeng, A.P., 2004a. Hierarchical structure and modules in the *Escherichia coli* transcriptional regulatory network revealed by a new top-down approach. BMC Bioinformatics 5, 199.

Ma, H.W., Kumar, B., Ditges, U., Gunzer, F., Buer, J., Zeng, A.P., 2004b. An extended transcriptional regulatory network of *Escherichia coli* and analysis of its hierarchical structure and network motifs. Nucleic Acids Res 32, 6643-6649.

Madan Babu, M., Teichmann, S.A., Aravind, L., 2006. Evolutionary dynamics of prokaryotic transcriptional regulatory networks. J Mol Biol 358, 614-633.

Martinez-Antonio, A., Collado-Vides, J., 2003. Identifying global regulators in transcriptional regulatory networks in bacteria. Curr Opin Microbiol 6, 482-489.

Milo, R., Shen-Orr, S., Itzkovitz, S., Kashtan, N., Chklovskii, D., Alon, U., 2002. Network motifs: simple building blocks of complex networks. Science 298, 824-827.

Moreno-Campuzano, S., Janga, S.C., Perez-Rueda, E., 2006. Identification and analysis of DNA-binding transcription factors in *Bacillus subtilis* and other Firmicutes--a genomic approach. BMC Genomics 7, 147.

Nijland, R., Kuipers, O.P., 2008. Optimization of protein secretion by *Bacillus subtilis*. Recent Pat Biotechnol 2, 79-87.

Pohl, S., Harwood, C.R., 2010. Heterologous protein secretion by *bacillus* species from the cradle to the grave. Adv Appl Microbiol 73, 1-25.

Price, M.N., Dehal, P.S., Arkin, A.P., 2007. Orthologous transcription factors in bacteria have different functions and regulate different genes. PLoS Comput Biol 3, 1739-1750.

Resendis-Antonio, O., Freyre-Gonzalez, J.A., Menchaca-Mendez, R., Gutierrez-Rios, R.M., Martinez-Antonio, A., Avila-Sanchez, C., Collado-Vides, J., 2005. Modular analysis of the transcriptional regulatory network of *E. coli*. Trends Genet 21, 16-20.

Rodriguez-Caso, C., Corominas-Murtra, B., Sole, R.V., 2009. On the basic computational structure of gene regulatory networks. Mol Biosyst 5, 1617-1629.

Sierro, N., Makita, Y., de Hoon, M., Nakai, K., 2008. DBTBS: a database of transcriptional regulation in *Bacillus subtilis* containing upstream intergenic conservation information. Nucleic Acids Res 36, D93-96.

Smits, W.K., Kuipers, O.P., Veening, J.W., 2006. Phenotypic variation in bacteria: the role of feedback regulation. Nat Rev Microbiol 4, 259-271.

Sonenshein, A.L., Hoch, J.A., Losick, R., 2002. *Bacillus subtilis*: From Cells to Genes and from Genes to Cells. In: Sonenshein, A.L., Hoch, J.A., Losick, R. (Eds.), *Bacillus subtilis* and Its Closest Relatives: From Genes to Cells. ASM Press, Washington, DC, pp. 3-5.

Song, C., Havlin, S., Makse, H.A., 2005. Self-similarity of complex networks. Nature 433, 392-395.

Thomas, R., 1998. Laws for the dynamics of regulatory networks. Int J Dev Biol 42, 479-485.

Thomas, R., Kaufman, M., 2001. Multistationarity, the basis of cell differentiation and memory. I. Structural conditions of multistationarity and other nontrivial behavior. Chaos 11, 170-179.

Vazquez, C.D., Freyre-Gonzalez, J.A., Gosset, G., Loza, J.A., Gutierrez-Rios, R.M., 2009. Identification of network topological units coordinating the global expression response to glucose in Bacillus subtilis and its comparison to Escherichia coli. BMC Microbiol 9, 176.
Wang, L., Perpich, J., Driks, A., Kroos, L., 2007. One perturbation of the mother cell gene regulatory network suppresses the effects of another during sporulation of *Bacillus subtilis*. J Bacteriol 189, 8467-8473.
Wojciechowski, M.F., Peterson, K.R., Love, P.E., 1991. Regulation of the SOS response in *Bacillus subtilis*: evidence for a LexA repressor homolog. J Bacteriol 173, 6489-6498.
Yan, K.K., Fang, G., Bhardwaj, N., Alexander, R.P., Gerstein, M., 2010. Comparing genomes to computer operating systems in terms of the topology and evolution of their regulatory control networks. Proc Natl Acad Sci U S A 107, 9186-9191.
Yu, H., Gerstein, M., 2006. Genomic analysis of the hierarchical structure of regulatory networks. Proc Natl Acad Sci U S A 103, 14724-14731.

# Figures

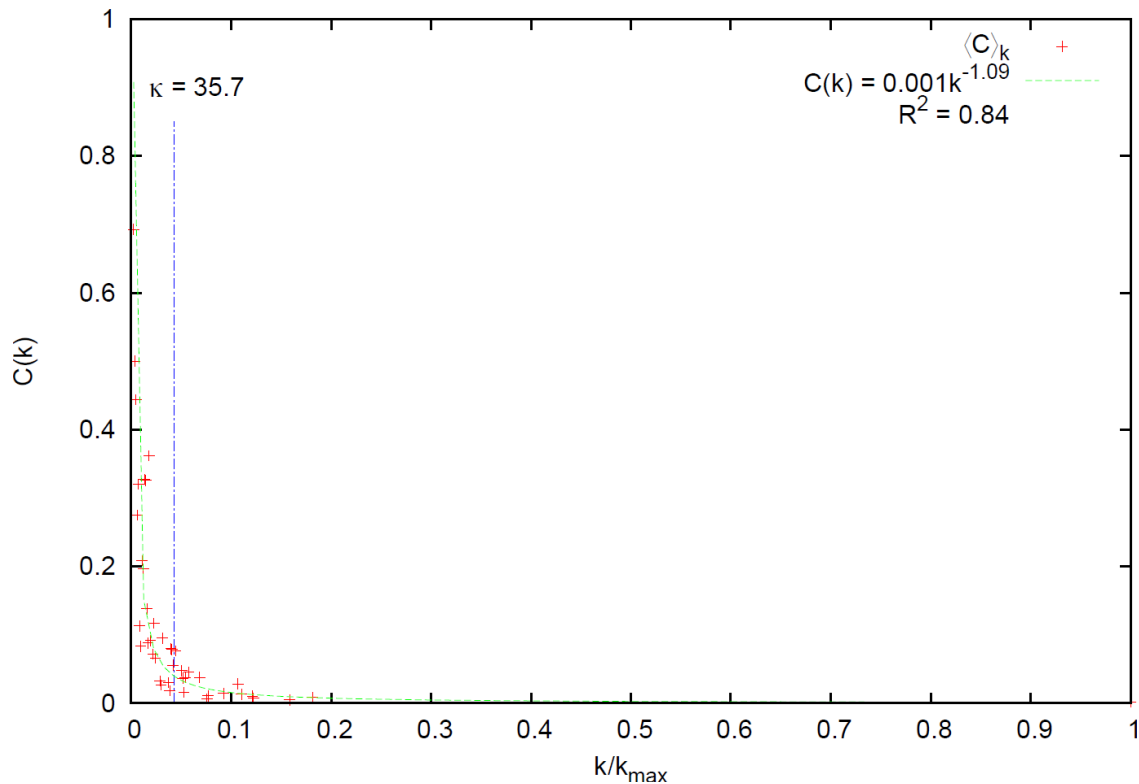

**Figure 1 – Identification of *B. subtilis* global TFs by using the κ-value**

The first step in the NDA is the identification of global TFs. We computed the κ-value for the *B. subtilis* TRN, which is indicated by a dot-dashed blue line. Global TFs are those genes with a connectivity greater that the κ-value. The green dashed line represents the distribution of the clustering coefficient, $C(k)$, used to compute the κ-value. Red crosses are the average clustering coefficient per connectivity.

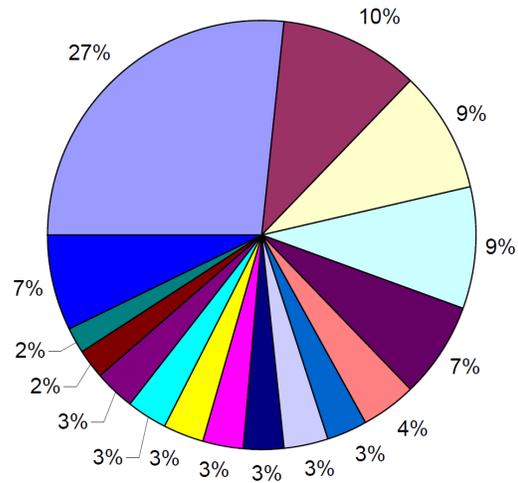

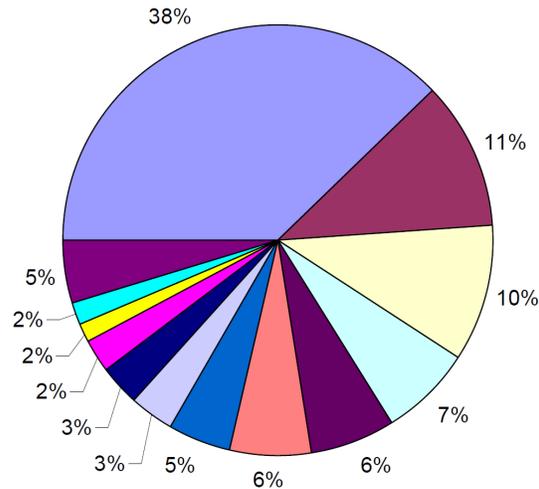

**Figure 2 – Distribution of modules per function for (a)** *B. subtilis* **and (b)** *E. coli.*

Percentage of modules devoted to a given class of physiological functions for each organism.

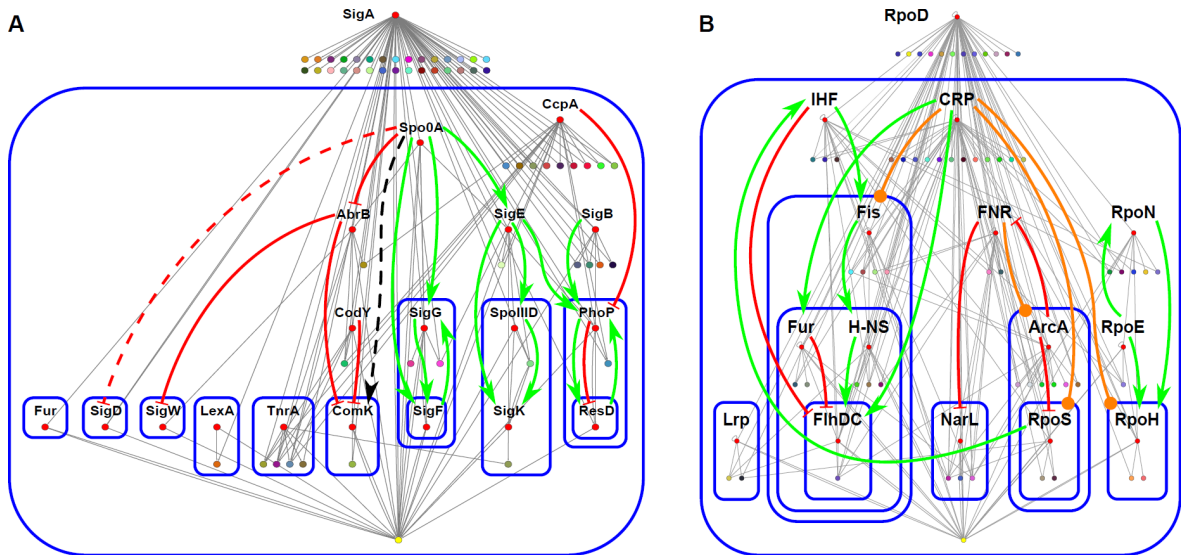

**Figure 3 – Functional architecture governing (a) *B. subtilis* and (b) *E. coli* TRNs**

Red nodes labeled with a name indicate global TFs. Nodes composing modules were shrunk into a single colored node. Continuous arrows (red for negative interactions, green for positive ones, orange for duals, and black for interactions whose sign is unknown) indicate regulatory interactions between global TFs. Dashed arrows represent a couple of recently curated interactions not included in our original dataset, which do not affect the results but provide more biological insights. Blue rounded-corner rectangles bound hierarchical layers. For the sake of clarity, SigA and RpoD interactions are not shown, and each megamodule is shown as a single yellow node at the bottom. SigA affects the transcription of all global TFs, except for CcpA, SpoIIID, SigW, LexA, TnrA, SigF and SigK. Additionally, SigA, SigB, SigD and SigH (the latter a modular TF) affect SigA expression. RpoD affects the transcription of all the global TFs, except for RpoE; while RpoD, RpoH and LexA (a modular TF) affect RpoD expression.

|  | E. coli | | | | |
|---|---|---|---|---|---|
| **B. subtilis** | **Global TFs** | **Modular genes** | **Intermodular genes** | **Strictly globally regulated genes** | **Total** |
| **Global TFs** | 3 (0.96%) | 4 (1.28%) | 0 | 1 (0.32%) | 8 (2.56%) |
| **Modular genes** | 3 (0.96%) | 106 (33.97%) | 23 (7.37%) | 34 (10.90%) | 166 (53.21%) |
| **Intermodular genes** | 0 | 0 | 3 (0.96%) | 0 | 3 (0.96%) |
| **Strictly globally regulated genes** | 0 | 73 (23.40%) | 13 (4.17%) | 49 (15.71%) | 135 (43.27%) |
| **Total** | 6 (1.92%) | 183 (58.65%) | 39 (12.50%) | 84 (26.92%) | 312 (100%) |

**Figure 4 – Conservation of systems-level components between *B. subtilis* and *E. coli***

By using TRN-wide orthologs, we analyzed the conservation between systems-level components for each organism. Cells color range from yellow to red and is proportional to the number of orthologs for a given pair of systems-level components.

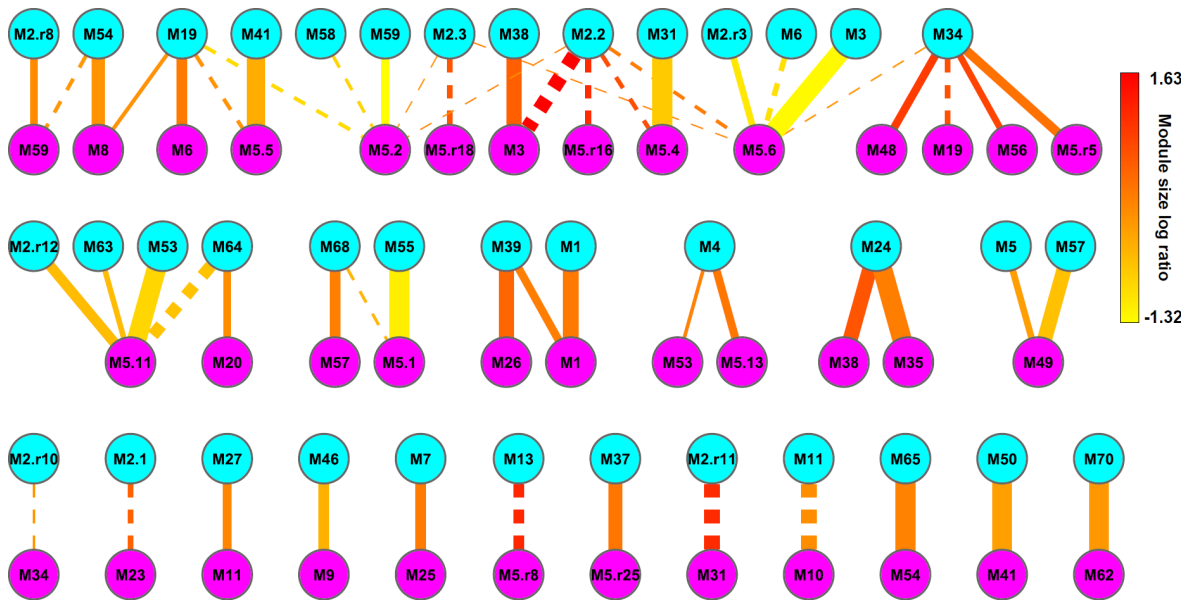

**Figure 5 – Module conservation between *B. subtilis* and *E. coli* via orthology**

*B. subtilis* modules are indicated by cyan nodes, while purple nodes represent *E. coli* modules. Line thickness is proportional to the module conservation index. Line color follows a gradient according to the module size log ratio thus indicating how modules differ in size. Dashed lines indicate pairs of modules that do not have equivalent functions.

Module pairs M70-M62 (DNA repair), M50-M41 (glycerol catabolism) and M65-M54 (biotin biosynthesis) exhibit the highest conservation in terms of module conservation index and size ratio.

# Tables

**Table 1 – Comparison between the systems-level composition of *B. subtilis* and *E. coli* TRNs**

|  | *B. subtilis* (1679 genes)[a] | *E. coli* (1692 genes)[a] |
|---|---|---|
| Global TFs | 19 (1.13%) | 15 (0.89%) |
| Modules | 69 + 1 megamodule (21 submodules) (45.74%) | 61 + 1 megamodule (39 submodules) (50.23%) |
| SGRGs | 850 (50.63%) | 691 (40.84%) |
| Intermodular genes | 42 (2.50%) | 136 (8.04%) |

[a]Percentages are relative to the total number of genes in each TRN.

**Table 2 – FBLs identified in the *B. subtilis* TRN**

| Type of FBL | Number of genes | Genes | Interactions |
|---|---|---|---|
| + | 2 | *comK rok* | – – |
| – | 2 | *gerE sigK* | – + |
| + | 2 | *phoP resD* | + + |
| – | 2 | *phoP resD* | – + |
| + | 2 | *rsfA sigG* | + + |
| + | 2 | *sigA sigB* | + + |
| + | 2 | *sigA sigD* | + + |
| + | 2 | *sigA sigH* | + + |
| + | 2 | *sigF sigG* | + + |
| – | 2 | *sigG spoVT* | + – |
| – | 3 | *abrB sigH sigA* | – + + |
| + | 3 | *abrB sigH spo0A* | – + – |
| + | 3 | *rsfA sigG sigF* | + + + |
| – | 3 | *sigF sigG spoVT* | + + – |
| + | 4 | *abrB sigH sigA spo0A* | – + + – |
| – | 4 | *rsfA sigG spoVT sigF* | + + – + |

**Table 3 – Functional conservation of modules composed by non-orthologous genes**

| *B. subtilis* modules | *E. coli* modules | Function |
|---|---|---|
| 48 | 40, 43 | Transport and catabolism of poliols (sugar alcohol) |
| 52 | 5.r7 | Transport and catabolism of 4-aminobutyrate |
| 61 | 45 | Divalent-metal-ion-dependent citrate transport |
| 2.r5, 42 | 36 | Transport and catabolism of lactose |
| 14 | 5.2, 46 | Periplasmic stress response |
| 11, 12 | 5.r12, 54 | Saturated and unsaturated fatty acid biosynthesis |
| 16 | 5.7 | Transport and catabolism of alpha-glucosides (maltose) |
| 2.4, 56 | 5.r9, 5.r10, 7 | Transport and catabolism of nucleotides/deoxyribonucleotides |
| 20, 21 | 19 | Phage-related functions |
| 44 | 5.9 | Transport and catabolism of hexuronates |
| 19, 68 | 4, 6, 9, 16, 17, 44, 58 | Transport and catabolism of pentoses |
| 2.1, 2.r1 | 5.12, 5.r20, 5.r22, 21, 41 | Anaerobiosis and fermentation |
| 2.1, 35 | 5.3 | Heat shock |
| 2.r13, 32, 58, 70 | 5.3, 34, 62 | DNA repair |
| 2.2, 2.3, 2.r16 | 33 | Extracytoplasmic function and stress response |

## Additional files

**Additional file 1 – Figure S1. Topological properties of the regulatory network of *B. subtilis***

Diverse topological properties of the transcriptional regulatory network of *B. subtilis* reconstructed for this study.

**Additional file 2 – Intermodular genes**

All the intermodular genes found in this study, their role as integrative elements and biological functions are discussed.

**Additional file 3 – Table S1. Module annotations**

All the modules identified in this study and their manual and computational annotations.

**Additional file 4 – Table S2. Conservation of modular functions**

Listing containing all pairs of modular functions conserved. Pairs of modules not conserving the biological function are highlighted in red. Highlighted in yellow are the pairs conserving the biological function but without statistical support (*p*-value > 0.05). Table is first sorted by ascending *p*-value. Then, rows that have the same *p*-value are further sorted by descending Module Conservation Index (MCI).

**Additional file 5 – Reconstructed regulatory network of *B. subtilis***

Flat file with the full data set of the *B. subtilis* transcriptional regulatory network reconstructed, as described in the Materials and methods section, for our analyses. Abbreviations in the flat file: TF, transcription factor; TG, target gene.